\documentclass{aastex63}

\newcommand{\lya}{Ly~$\alpha$}
\newcommand{\lyb}{Ly~$\beta$}
\newcommand{\lyg}{Ly~$\gamma$}

\newcommand{\photu}{photon units }
\newcommand{\galex}{{\it GALEX }}
\newcommand{\voyager}{{\it Voyager }}

\shorttitle{The Diffuse UV/Optical Background}
\shortauthors{Murthy et al.}
\graphicspath{{./}{figures/}}

\begin{document}

\title{The Diffuse Ultraviolet and Optical Background: Status and Future Prospects}

\author{Jayant Murthy}
\email{jmurthy@yahoo.com}
\affiliation{Indian Institute of Astrophysics, Bengaluru 560 034, India \\}

\author{M. S. Akshaya}
\affiliation{CHRIST (Deemed to be University), Bengaluru 560 029, India \\}

\author{S. Ravichandran}
\affiliation{CHRIST (Deemed to be University), Bengaluru 560 029, India \\}



\begin{abstract}

The ultraviolet and optical background forms a baseline for any observation of the sky. It includes emission lines and scattered light from the atmosphere; resonant scattering from the Lyman lines of interplanetary hydrogen and the scattering of sunlight from Solar System dust (zodiacal light); scattering of starlight from interstellar dust (DGL) with emission from molecular hydrogen fluorescence or from line emission in selected areas; and an extragalactic component seen most easily at high Galactic latitudes. We will discuss the different components of the diffuse radiation field in the UV and the optical. We close with a hope that there will be new observations from missions near the edge of the Solar System.

\end{abstract}



\section{Introduction} \label{sec:intro}

\begin{figure}
\begin{center}
 \includegraphics[width=0.8\textwidth]{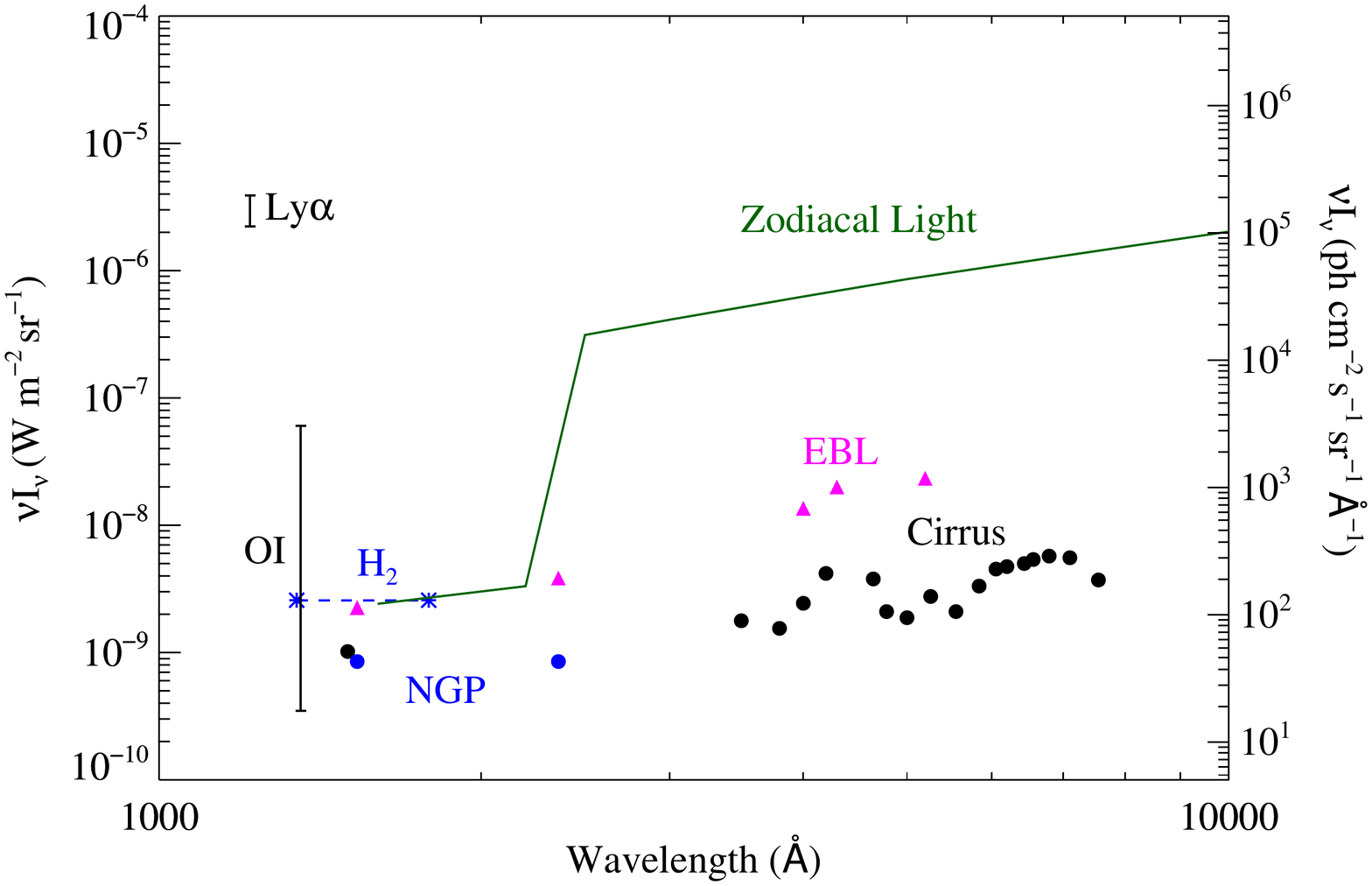} 
 \caption{Contributors to the diffuse sky where the airglow lines are from \citet{Brune1978}, \citet{Meier1991}, and \citet{Leinert1998}; zodiacal light from \citet{Leinert1998}; molecular hydrogen fluorescence in the Lyman and Werner bands from \citet{Jo2017} and \citet{Akshaya2018}; cirrus \citep{Mattila1979, Mattila1990, Haikala1995, Akshaya2018}; and extragalactic background light \citep{Mattila2017, Akshaya2018}.}
  \label{Fig:diffuse_contributors}
\end{center}
\end{figure}

Observations of the sky background before the space age were dominated in most spectral regions by atmospheric emission and this was reflected in the name and composition of the first organized group to look at the diffuse sky: \textit{The Light of the Night Sky} (IAU Commission 21). The first meetings of the new Commission focused on atmospheric emission and most of the attendees were atmospheric scientists \citep{MurthyIAU}. The Commission played an active role in fostering international collaborations and results from that period played an important role in our modern understanding of the atmosphere \citep{Meier1991}. This situation changed rapidly as atmospheric scientists moved to the larger geophysical community and a new group of astronomers interested in the sky background from Galactic and extragalactic sources over the entire spectrum entered the field. However, it was recognized that all components of the night sky were present in every observation and in all spectral regions and thus the Commission maintained an eclectic membership until the recent IAU reorganization. Amongst the many significant contributions from IAU Commission 21 was the review of the diffuse light by \cite{Leinert1998} which brought together results from many different disciplines, each with its own set of units, into a single reference summarizing the wide range of contributors over the entire electromagnetic spectrum. Observations of the low surface brightness universe must take care to account for all components of the dark sky (Fig. \ref{Fig:diffuse_contributors}).

The first observations of the diffuse background at any wavelength were made by accident when \cite{Giacconi1962} observed that the Moon was occulting the diffuse X-ray background. This was closely followed by the discovery of the cosmic microwave background \citep{Penzias1965} and then the ultraviolet background  \citep{Hayakawa1969, Lillie1976}. It was soon realized that the sky is extraordinarily dark in the UV with a sky brightness as low as 26 mag (arcsec)$^{-2}$ (see \citet{oconnell1987} for details and definitions). There were therefore a series of observations, many motivated by a search for an extragalactic background, but with disagreement over both the measurements and the interpretation \citep{Bowyer1991, Henry1991}. Observations in the visible were even more difficult because of the zodiacal light and the first reliable measurements of the visible sky background and the much fainter extragalactic background came from {\it Pioneer 10/11} observations \citep{Toller_EBL}.

There have been considerable advances since those early days \citep{Murthy2009} and we will describe the observations and the current  interpretation of the diffuse Galactic light (DGL) and the extragalactic background light (EBL) in the UV and optical. The contributors and their magnitudes are summarized in Fig. \ref{Fig:diffuse_contributors}. The darkness of the UV sky is obvious with few contributors other than \lya\ and possibly the O I lines (1304/1356 \AA). The flux from the Sun increases dramatically at wavelengths greater than 3000 \AA\ and it is very difficult to extract any of the components of the diffuse sky from the bright zodiacal light. We will begin by defining the spectral regions and the terms reviewed in this work, go through the different contributors, discuss the observations from whence the data come from, and finally summarize the results. We note that the comprehensive survey by \citet{Leinert1998} is still the starting point for any study of the diffuse sky.

\section{Overview}
\subsection{Spectral Bands}

The ultraviolet and optical bands stretch from 100 \AA\ to 10,000 \AA\ (10 nm to 1 \micron) but we will spend most of this review on the ultraviolet (100 -- 3200 \AA) where most of the observations are. Detectors in the UV are typically solar-blind, intensified detectors with either CCDs or with electronic readouts (see the review of UV detectors by \cite{Joseph1995}). They are close to zero-noise and photon counting which allows for long integration times and detections limited only by the photon noise. This is different from the visible where instrumental noise and the foreground from atmospheric and interplanetary emission are both brighter (Fig. \ref{Fig:diffuse_contributors}) and are the primary limiting factors. We have listed the different bands below:
\begin{itemize}
    \item FUV (912 -- 1200 \AA): This wavelength range is bracketed by the interstellar Lyman limit (912 \AA) below which there will be no astrophysical radiation and the intense \lya\ line (1216 \AA) on the long wavelength side.  The detectors must be windowless which complicates their handling but, more importantly, this spectral region is dominated by atmospheric and interplanetary \lyb\ (1026 \AA) and \lya\ (1216 \AA). Although efforts were made to depress the \lya\ count rate (eg. \citet{Stern_NH_2008}), the scattering wings of \lya\ can stretch through the entire spectral region.
    \item FUV (1250 -- 1800 \AA): The intense \lya\ line can be excluded using any one of a number of filters, such as CaF$_{2}$ (calcium fluoride), leaving only the atmospheric O I lines (1304 and 1356 \AA), which largely vanish at night. The long wavelength cutoff may be achieved using a CsI (cesium iodide) photocathode with a long-wavelength cutoff at about 1800 \AA. Note that, by convention, this band is also usually called the FUV.
    \item NUV (1600 -- 3000 \AA): Instruments in this wavelength range use a CsTe (cesium telluride) or RbTe (rubidium telluride) photocathodes, either of which is solar-blind with a long wavelength cutoff of about 3000 \AA, with a filter such as sapphire (1450 \AA) or quartz (1600 \AA) to set the lower limit. There are few airglow lines in this range in the night sky with O I at 2471 \AA\ being the brightest. Zodiacal light becomes important with the increase in the Solar spectrum at 2000 \AA\ and begins to dominate the flux at longer wavelengths, depending on the helioecliptic latitude and longitude (as measured from the Sun) and the Galactic latitude.
    \item VIS (3000 -- 10,000 \AA): There is much more background in the visible, both from the night sky and from the zodiacal light. As a G star, the Sun is much brighter in the visible than in the UV and any background measurement will be dominated by a large zodiacal light component \citep{Leinert1998} and, if observed from the ground, by the sky brightness \citep{Mattila2017}. The best measurements of the sky background in the optical have therefore come from spacecraft near the edge of the Solar System \citep{Toller_EBL}.
\end{itemize}

\subsection{Contributors to the Diffuse Background}

\begin{figure}[t]
\begin{center}
 \includegraphics[width=0.8\textwidth]{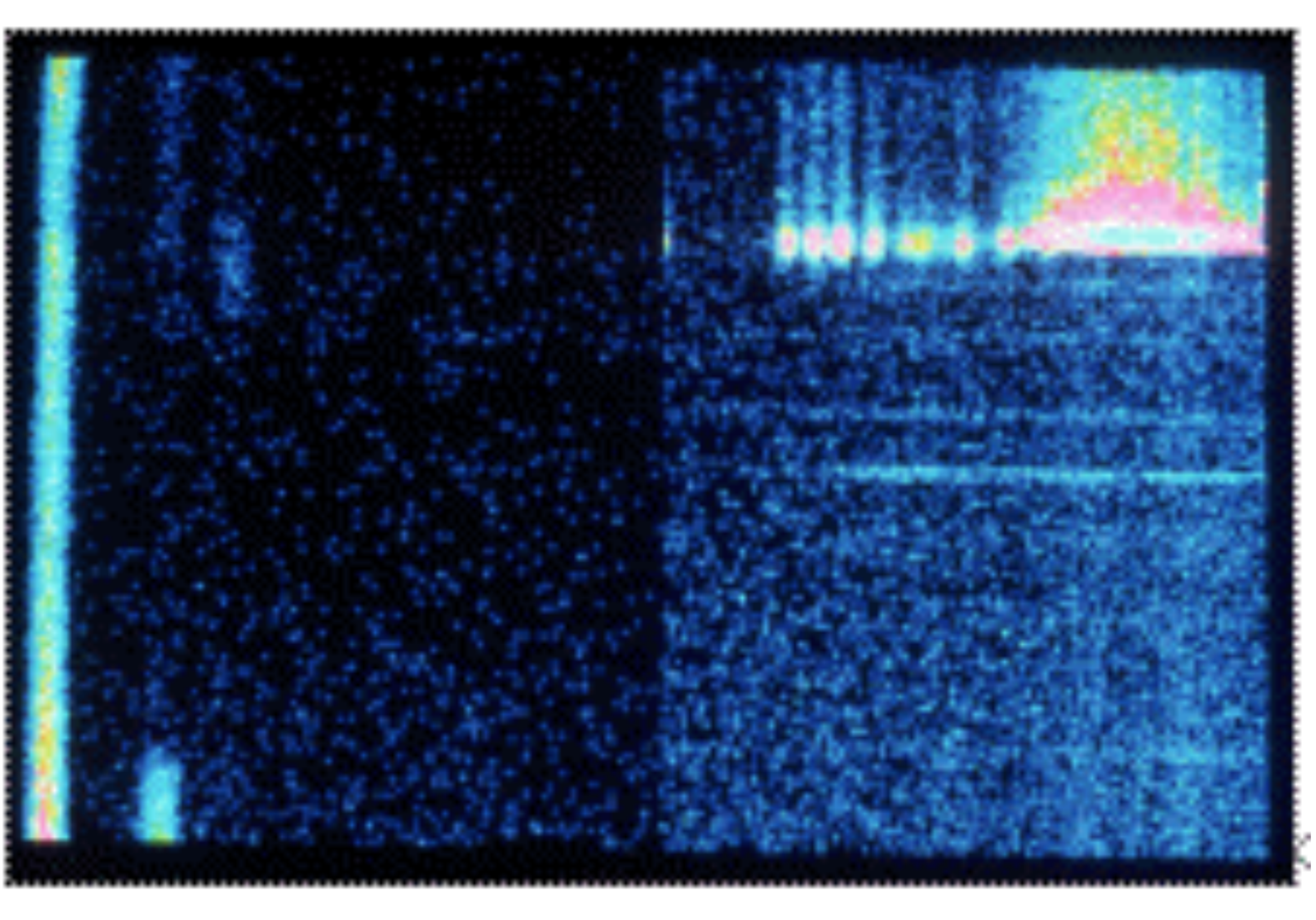} 
 \caption{Observations of the UV sky background from the {\it Space Shuttle} \citep{murthyUVX}.}
  \label{Fig:UVX}
\end{center}
\end{figure}

All the contributors to the diffuse background are shown in Fig. \ref{Fig:UVX} which is a 20 minute observation of a target at a low Galactic latitude made with the UVX (ultraviolet experiment) on board the Space Shuttle Columbia (STS-61C) \citep{murthyUVX}. There were two spectrographs on board, one in the FUV (1200 -- 1700 \AA) and the other in the NUV (1700 -- 3000 \AA) and individual photon hits from both instruments have been plotted as a function of wavelength (x-axis) and time (y-axis), with dusk being at the bottom and dawn at the top. There was much more noise in the NUV instrument because of a faulty wire and this is easily seen as the much brighter background in the right half of the plot. Most of the remaining photons are from the sky. The nighttime sky is much darker with emission limited to the Lyman lines of hydrogen at 1026 and 1216 \AA\ and the O I lines at 1304 and 1356 \AA. As seen in Fig. \ref{Fig:UVX}, \lya\ is strong throughout the observation while the oxygen lines vanish a few minutes after dusk and only reappear a few minutes before dawn. The bright band in the dawn in the NUV spectrograph is due to nitric oxide (NO).

There are two other features worth noting in the figure. The first is the rise at long wavelengths due to zodiacal light: sunlight scattered from interplanetary dust grains. The other is the several horizontal lines in the NUV part of the spectrum which is due to A stars passing through the field of view. Because they are relatively cool stars, they do not show up in the FUV.

\subsection{Observations}

\begin{longtable}{lllp{9cm}}
\caption{Diffuse Observations} \label{TAB:missions} \\
\hline
Reference & Wavelength & Offset & Notes  \\
\hline
\citet{Kurt1968} & 1225 -- 1340 & - & Photometers on Venus 2 and 3. \\
\citet{Roach1968} & 5300 -- 6300 & - & Zenith photometers at mid-latitude stations.\\
\citet{Hayakawa1969} & 1350 -- 1480 & - & Rocket-borne Geiger-M\"{u}ller counters.\\
\citet{Lillie1969} & 2100 -- 2800 & - & Rocket-borne photometers. \\ 
\citet{Lillie1972} & 1250 -- 4300 & - & Surface brightness of night sky using stellar photometers on \textit{OAO-2}. \\
\citet{Witt1973} & 1500 -- 4200 & - & \textit{OAO-2} observations in 29 of Kapteyn's selected areas. \\
\citet{Lillie1976} & 1500 -- 4200 & - & \textit{OAO-2} observations between Galactic longitudes $65^{\circ} < l < 145^{\circ}$. \\
\citet{Mattila1976} & 3450 -- 4700 & - & Measurements from \textit{Kitt Peak National Observatory} and \textit{Steward Observatory}. \\
\citet{Henry1977ApJ} & 1425 -- 1640 & - & Geiger counter carried by an Aerobee rocket. \\
\citet{Dube1977} & 5100 & - & Observations from \textit{Cerro Tololo Inter-American Observatory} and \textit{Kitt Peak National Observatory}. \\
\citet{Morgan1978} & 1550 -- 2740 & - & Observations with \textit{S2/68} sky-survey telescope on the \textit{TD-1} satellite. \\
\citet{Henry_dust1978} & 1180 -- 1680 & - & \textit{Apollo 17} spectroscopic observations of two regions at moderate Galactic latitudes. \\
\citet{Henry_ngp1978} & 1180 -- 1680 & - & \textit{Apollo 17} spectroscopic observations of the Galactic Pole regions. \\
\citet{Spinrad1978} & 3960 -- 4700 & - & Observations from \textit{Lick Observatory}. \\
\citet{Anderson1979} & 1230 -- 1680 & $285 \pm 32$ & Spectroscopic observations at high latitudes from an \textit{Aries} rocket. \\
\citet{Mattila1979} & 3450 -- 7550 & - & Observations of Lynds 134 and Lynds 1778/1780 from \textit{Kitt Peak National Observatory}, \textit{Steward Observatory} and \textit{ESO}. \\
\citet{Dube1979} & 5115 & - & Observations from \textit{Kitt Peak National Observatory}. \\
\citet{Paresce1980} & 1350 -- 1550 & $<300$ & Observations from \textit{EUV} telescope on the \textit{Apollo-Soyuz} mission. \\
\citet{Feldman_hotgas1981} & 1200 -- 1670 & $150 \pm 50$ & Spectroscopic observations near the NGP from an \textit{Aries} rocket. \\
\citet{Zvereva1982} & 1100 -- 1850 & - & Photoelectric spectrometer (\textit{Galaktika}) on board \textit{Prognoz-6} satellite. \\
\citet{Joubert1983} & 1690 & 300 -- 690 & High latitude survey with \textit{ELZ} spectrometer on board the \textit{D2B-Aura} Satellite. \\
 & 2200 & 160 -- 360 & \\
 \citet{Weller1983} & 1220 -- 1500 & - & FUV observations from \textit{Solrad 11} satellite. \\
 \citet{Toller_EBL} & 4400 & - & Background light observations using photopolarimeter on board the \textit{Pioneer 10} spacecraft. \\
 \cite{Jakobsen1984} & 1590 & $<550$ & Broad band photometers on board an \textit{Aries} sounding rocket. \\
  & 1710 & $<900$ & \\
  & 2135 & $<1300$ & \\
\citet{Holberg1986} & 500 -- 1200 & 100 -- 200 & Spectrum near NGP from \textit{Voyager 2}. \\
\citet{Boughn1986} & 6500 & - & Surface photometry of L134 using \textit{Kitt Peak National Observatory}. \\
\citet{Tennyson1988} & 1700 -- 2850 & $300 \pm 100$ & Spectroscopic measurements at high Galactic latitude from an \textit{Aries} rocket. \\
\citet{FixJ1989} & 1500 & $530 \pm 80$ & Imaging instrument on the \textit{Dynamic Explorer I} satellite. \\
\citet{murthyUVX} & 1200 -- 1700 & - & Spectra of diffuse FUV background from Johns Hopkins \textit{UVX} experiment aboard Space Shuttle \textit{Columbia (STS-61C)}. \\
\citet{Murthy1990} & 1650 -- 3100 & - & Spectra from the Johns Hopkins \textit{UVX} experiment. \\
\citet{Mattila1990} & 3500 -- 5550 & - & Observations of L1642 using \textit{European Southern Observatory} telescope. \\
\citet{Murthy1991} & 700 -- 1100 & - & \textit{Voyager 2} observations of various Galactic latitudes \\
\citet{Hurwitz1991} & 1415 -- 1835 & 50 & BEST Spectrometer on the Space Shuttle. \\
\citet{Onaka1991} & 1560 & 200 -- 300 & Imager aboard the sounding rocket S520-8. \\
\citet{Henry1993} & 1500 & $300 \pm 100$ & \textit{UVX} observations above $|b|=40^{\circ}$ from the Space Shuttle. \\
\citet{Murthyvoy1993} & 912 -- 1150 & - & \textit{Voyager} observations of the Eridanus superbubble. \\
\citet{Murthy_CS1994} & 912 -- 1600 & - & \textit{Voyager} observations of the Coalsack nebula. \\
\citet{Hurwitz1994} & 1400 -- 1850 & - & BEST spectra of the Taurus molecular cloud from the Space Shuttle (STS-61C). \\
\citet{Gordon1994} & 1230 -- 2000 & - & Imaging observations of Upper Scorpius from \textit{NRL-803} on \textit{STS-39}. \\
\citet{Haikala1995} & 1400 -- 1800 & - & \textit{Far Ultraviolet Space Telescope (FAUST)} observations of Galactic cirrus cloud G251.2+73.3. \\
\citet{Waller1995} & 1500 -- 2500 & - & Images from Shuttle-borne \textit{Ultraviolet Imaging Telescope (UIT)}. \\
\citet{Leinert1995} & 3500 -- 8200 & - & Measurement of sky brightness from \textit{Calar Alto Observatory}. \\
\citet{Sasseen1996} & 1400 -- 1800 & - & \textit{FAUST} observations at different Galactic latitudes. \\
\citet{Mattila1996} & 3500 -- 5470 & - & Measurement of sky brightness from \textit{ESO La Silla Observatory}. \\
\citet{Murthy1997} & 1420 -- 2670 & - & Galactic plane observations from \textit{Midcourse Space Experiment (MSX)} satellite. \\
\citet{Witt1997} & 1400 -- 1800 & $160 \pm 50$ & \textit{FAUST} observations. \\
\citet{Murthy_voy} & 912 -- 1100 & $<$30 & All-sky \textit{Voyager} \textit{UVS} observations. \\
\citet{Murthy2001} & 1400 -- 2600 & - & \textit{MSX} observations of diffuse emission in Orion. \\
\citet{Schiminovich2001} & 1740 & $200 \pm 100$ & {\it NUVIEWS} rocket experiment in NUV. \\
\citet{Bernstein2002} & 3000 -- 8000 & - & \textit{HST} and ground-based observations from \textit{Las Campanas Observatory}. \\
\citet{Gibson2003} & 1650 -- 2200 & - & Observations of Pleiades reflection nebulosity using \textit{Wide-Field Imaging Survey Polarimeter (WISP)}. \\
\citet{Murthy_sahnow2004} & 1000 -- 1200 & - & \textit{Far Ultraviolet Spectroscopic Explorer (FUSE)} observations. \\
\citet{Sujatha2005} & 1100 & - & \textit{Voyager} observations of Ophiuchus. \\
\citet{Murthy_Orion2005} & 900 -- 1200 & - & \textit{FUSE} observations of Orion nebula (M42). \\
\citet{Lee2006}  & 1370 -- 1670 & - & Observations of the Taurus molecular cloud using {\it Spectroscopy of Plasma Evolution from Astrophysical Radiation (SPEAR/FIMS)} flown aboard \textit{STSAT-1} satellite. \\
\citet{Edelstein2006} & 1360 -- 1710 & - &  Spectroscopic map from \textit{SPEAR\ /FIMS}. \\  
\citet{Sujatha2007} & 900 -- 1200 & - & \textit{FUSE} and \textit{Voyager} observations of Coalsack Nebula. \\
\citet{Lee2008} & 1370 -- 1670 & - & Observations of Ophiuchus molecular cloud region using \textit{SPEAR} imaging spectrograph. \\
\citet{Sujatha2009} & 1350 -- 2850 & - & {\it Galaxy Evolution Explorer (GALEX)} observations of the Sandage cloud. \\
\cite{Puthiyaveettil2010} & 1400 -- 1900 & 500 & Observations from the \textit{Dynamics Explorer spacecraft, DE-1}. \\
\citet{Sujatha2010} & 1350 -- 1750 & $30 \pm 10$ & \textit{GALEX} observations of Draco. \\
 & 1750 -- 2850 & $49 \pm 13$ & \\
 \citet{Murthy_galex_data2010} & 1530 & - & Diffuse UV cosmic background using \textit{GALEX} archival data. \\
  & 2310 & - & \\
 \citet{Seon2011} & 1370 -- 1710 & $\approx$300 & Observations from \textit{SPEAR}. \\
 \citet{Matsuoka2011} & 4400 -- 6400 & - & Measurements of cosmic optical background from \textit{Pioneer 10/11 Imaging Photopolarimeter (IPP)} data. \\
 \citet{Mattila2012} & 4000 -- 5200 & - & Spectroscopic measuurements of EBL using \textit{ESO's VLT} telescope. \\ 
 \citet{Murthyvoy_all} & 500 -- 1600 & - & \textit{Voyager} observations of diffuse FUV radiation field. \\
 \citet{Jo2012} & 1350 -- 1750 & - & Observations from \textit{Far-Ultraviolet Imaging Spectrograph (FIMS)} on board Korean microsatellite \textit{STSAT-1}. \\
 \citet{Hamden2013} & 1344 -- 1786 & 300 & All-sky map from \textit{GALEX} observations. \\
 \citet{Lim2013} & 1350 -- 1750 & - & Observations of the Taurus-Auriga-Perseus complex from \textit{GALEX} and \textit{FIMS}. \\
\citet{Boissier2015} & 1528 & 315 & \textit{GALEX} ultraviolet Virgo Cluster Survey (\textit{GUViCS}). \\
 \citet{Jyothy2015} & 1571 & -  & \textit{GALEX} observations near the Aquila Rift. \\ 
  & 2317 & - & \\
 \citet{Murthy2016} & 1531 & 300 & All-sky observations of diffuse background from \textit{GALEX}. \\
 & 2361 & 600 & \\
 \cite{Mattila2017II} & 3800 -- 5800 & - & Surface brightness measurement of L1642 using \textit{VLT/FORS} at \textit{ESO}. \\
 \citet{Akshaya2018} & 1539 & 120 -- 180 & \textit{GALEX} observations of the North and South Galactic Poles. \\
  & 2316 & 300 -- 400 & \\
 \citet{Akshaya2019} & 1539 & $240 \pm 18$ & \textit{GALEX} observations of Galactic latitudes between $70^{\circ}<b<80^{\circ}$. \\
  & 2316 & $394 \pm 37$ & \\
\hline
\end{longtable}

Barring a few rocket experiments \citep{Bowyer1991, Henry1991}, most observations of the diffuse background have come from instruments and missions intended for other purposes and which have been re-purposed for diffuse observations. We have compiled a list of these in Table \ref{TAB:missions}. Most of the diffuse observations in the wavelength range from 900 -- 1100 \AA\ have come from the \voyager observations taken during their cruise phase \citep{Murthy_voy, Murthyvoy_all} with all-sky observations in the 1300 -- 3000 \AA\ coming from \galex \citep{Murthy_galex_data2010, Hamden2013, Murthy2016} and {\it SPEAR} \citep{Edelstein2006, Seon2011}. Observations in the visible have been even more difficult because of the bright foregrounds and the best have come from the Pioneer spacecraft \citep{Toller_EBL, Gordon_Pioneer1998, Matsuoka2011} and the LORRI instrument on New Horizons \citep{Zemcov_lorri_2017}.

\section{Sources}

The diffuse radiation field can be divided into several components which may all be important in different spectral regions and at different locations in the sky.

\subsection{Atmospheric Emission}

Of course, the diffuse background can only be observed at night from any ground or space-based telescope and the brightest lines in the UV are the resonance lines of \lyg\ (973 \AA), \lyb\ (1026 \AA) and \lya\ (1216 \AA). The \lya\ line, in particular, may be bright enough that its scattering wings contaminate the entire spectrum. The O I lines of 1304 \AA, 1356 \AA\ and 2471 \AA\ may contribute close to the terminator but are small in the orbital night. In principle, ground-based observations are possible in the visible but, even at night, it is very difficult to extract the astrophysical emission from the atmospheric foreground \citep{Mattila_EBL2019}.

\subsection{Interplanetary Emission}

The interplanetary emission may be divided into two regimes. The UV includes line emission from interplanetary hydrogen at \lyb\ and \lya\ (and helium at 584 \AA) due to the resonant scattering of Solar photons. There is a negligible amount of solar radiation in the UV \citep{Mount_sun1985} and the scattering of the solar continuum only becomes important at wavelengths greater than 2500 \AA\ and dominates the diffuse sky in the visible (Fig. \ref{Fig:diffuse_contributors}). Even in the UV, scattering from the interplanetary lines may contaminate the entire spectrum \citep{Murthyvoy_all} and observations  made from within the Solar System will always have to deal with a bright and perhaps complex interplanetary foreground and the best limits have come from spacecraft far away from the Sun, where the interplanetary emission is much reduced \citep{Zemcov_lorri_2017}.

\subsection{Dust Scattering}

\begin{figure}[t]
\begin{center}
 \includegraphics[width=0.45\textwidth]{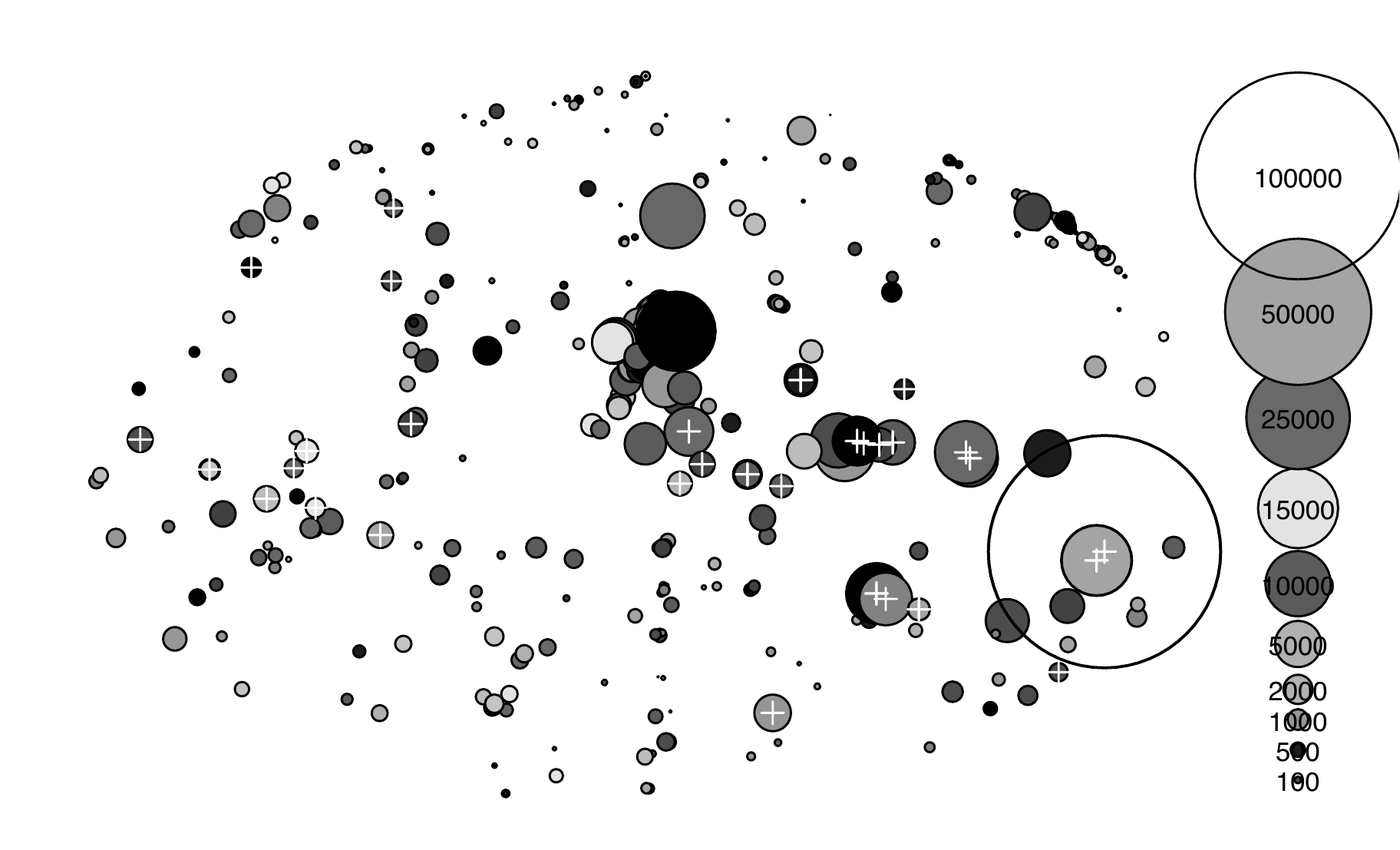} 
 \includegraphics[width=0.45\textwidth]{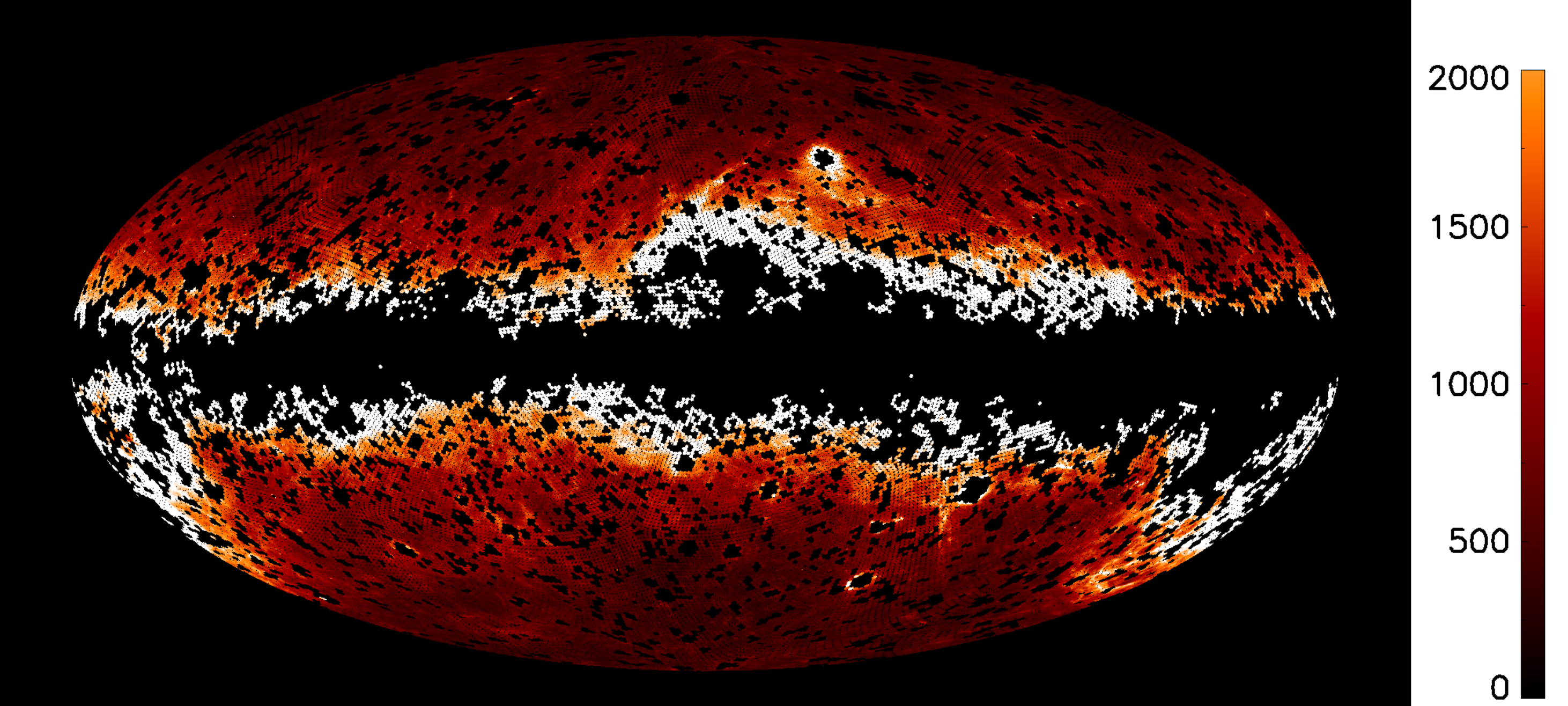} 
 \includegraphics[width=0.45\textwidth]{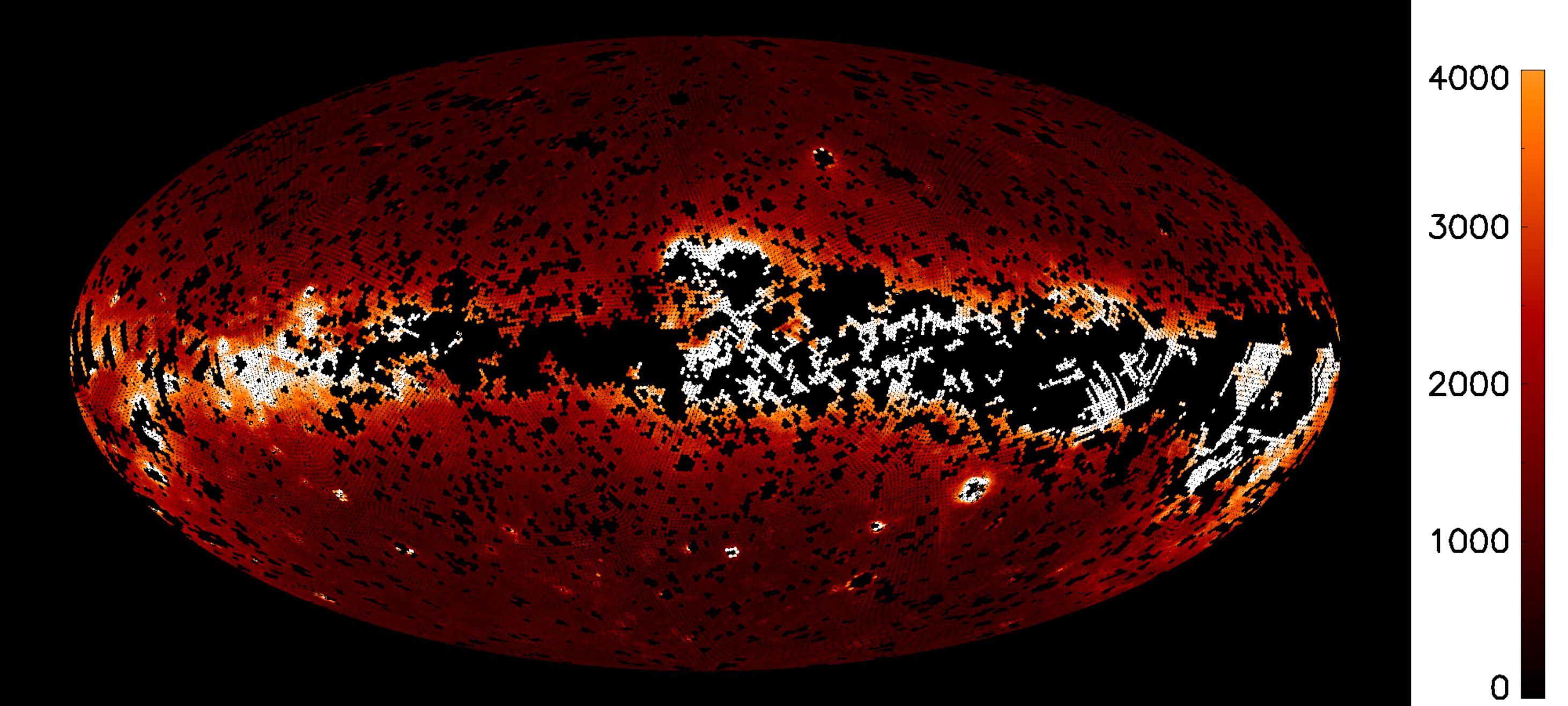} 
 \includegraphics[width=0.45\textwidth]{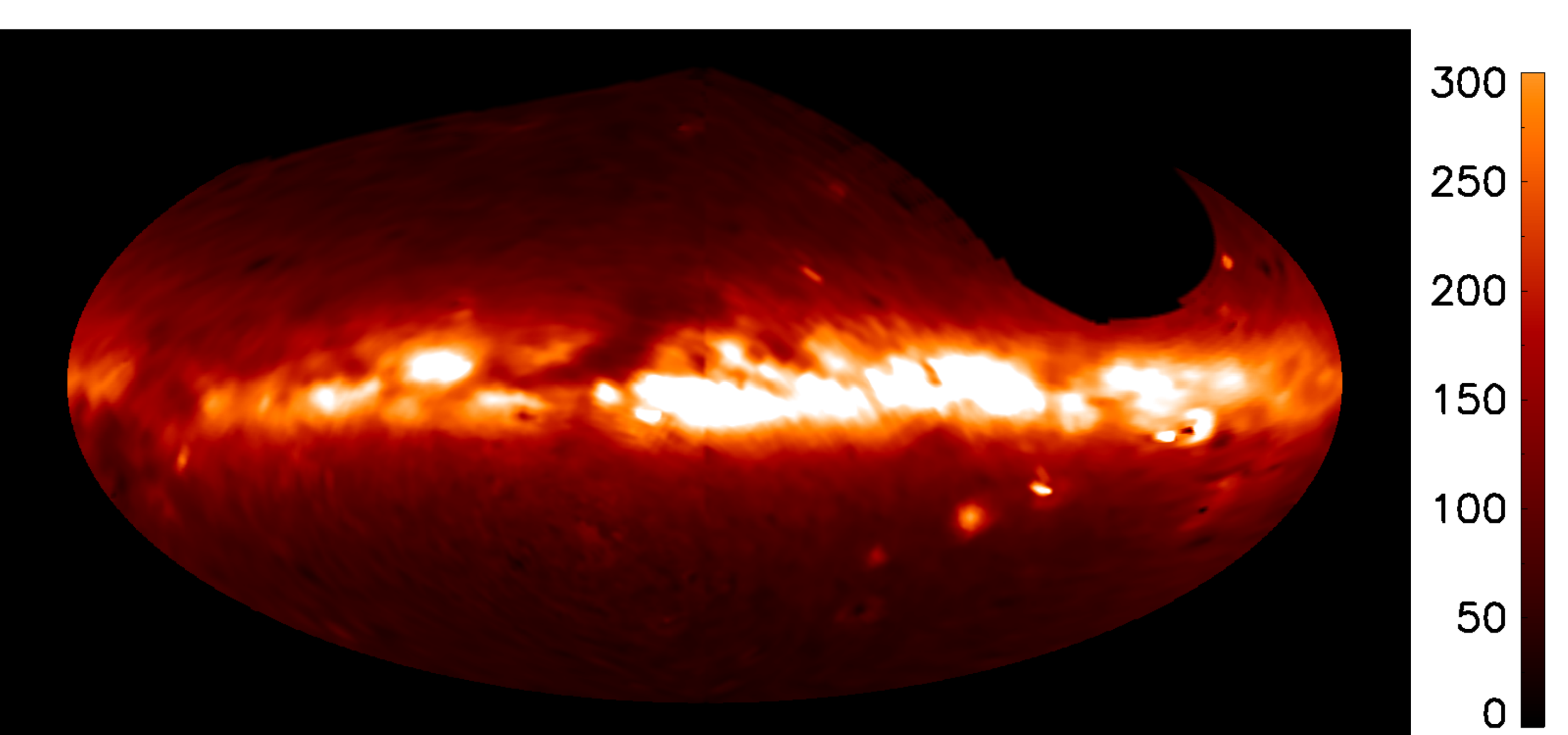} 
 \includegraphics[width=0.45\textwidth]{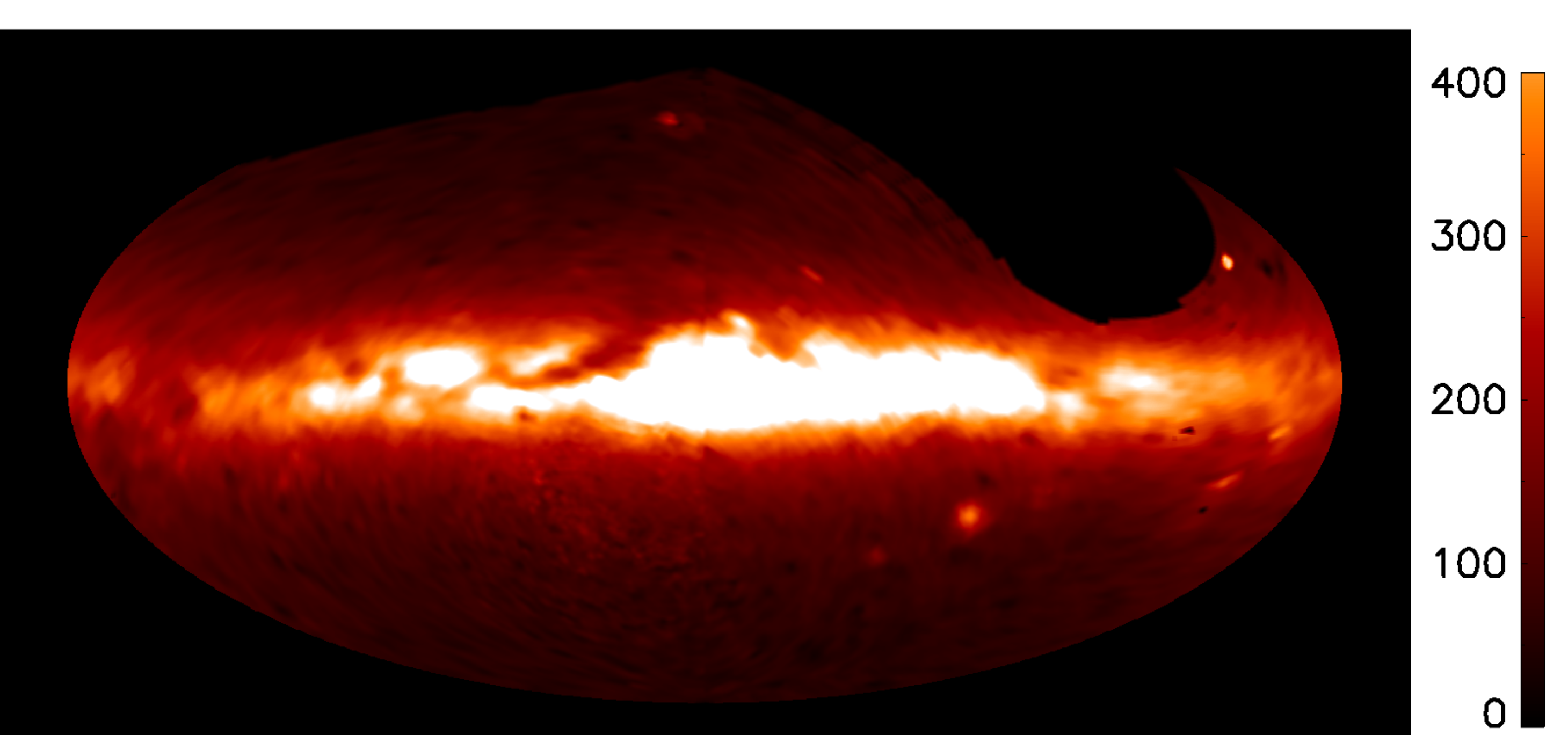} 
\caption{Clockwise from top left: observations of the UV sky background at 1100 \AA\ from the \voyager UVS \citep{Murthyvoy_all} supplemented by 32 observations from the {\it FUSE} mission \citep{Murthy_sahnow2004}, marked by crosses; observations from the {\it GALEX} FUV and NUV \citep{Murthy2014apj}; and observations from the {\it Pioneer 10/11} at 4370 and 6441 \AA\ \citep{Gordon_Pioneer1998}. All plots are in Galactic coordinates with the origin (gl = 0; gb = 0) in the center of the plot. }
  \label{Fig:euv_bkgd}
\end{center}
\end{figure}

\begin{figure}[t]
\begin{center}
 \includegraphics[width=0.45\textwidth]{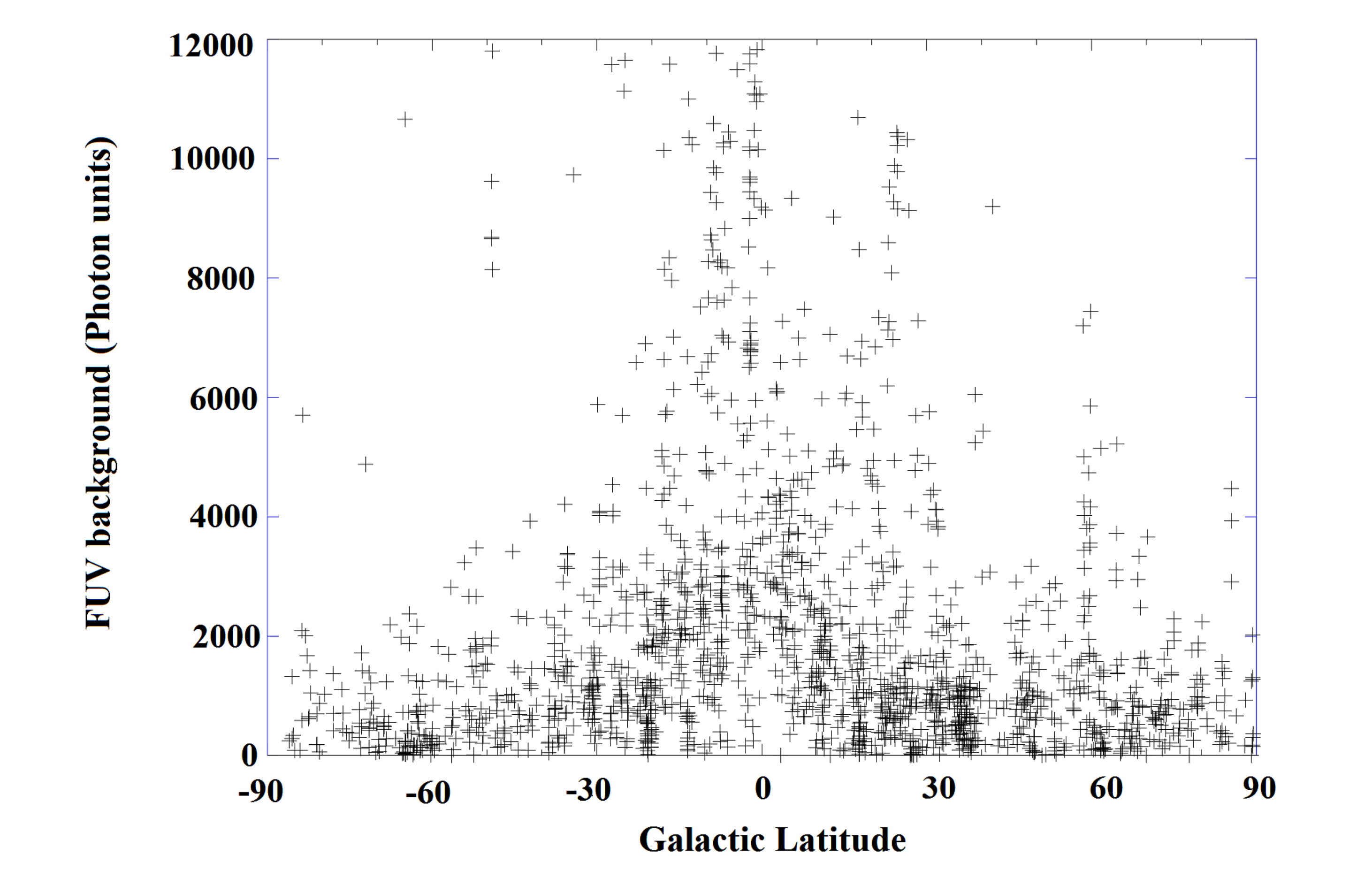} 
 \includegraphics[width=0.45\textwidth]{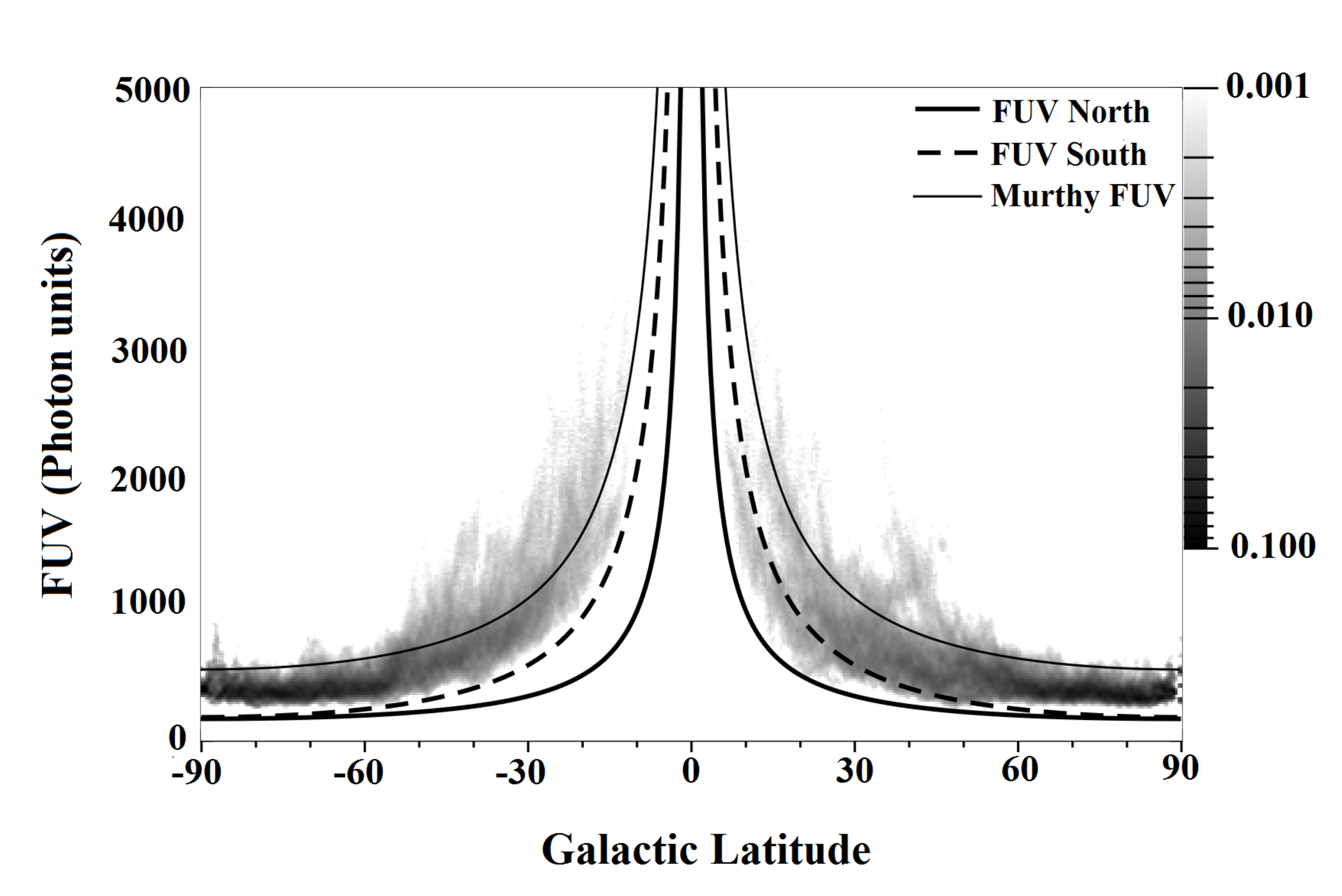} 
 \includegraphics[width=0.4\textwidth]{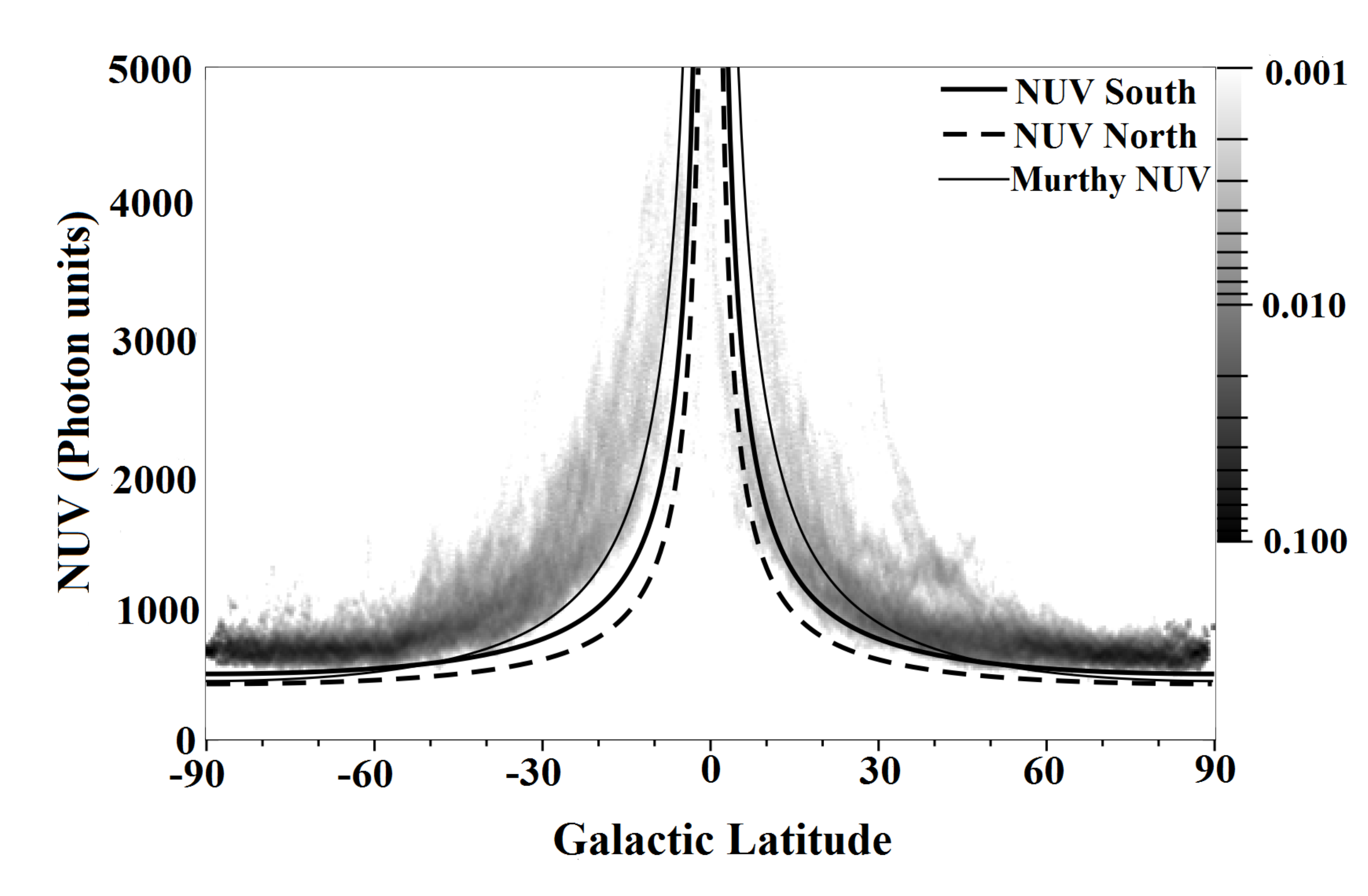} 
 \includegraphics[width=0.45\textwidth]{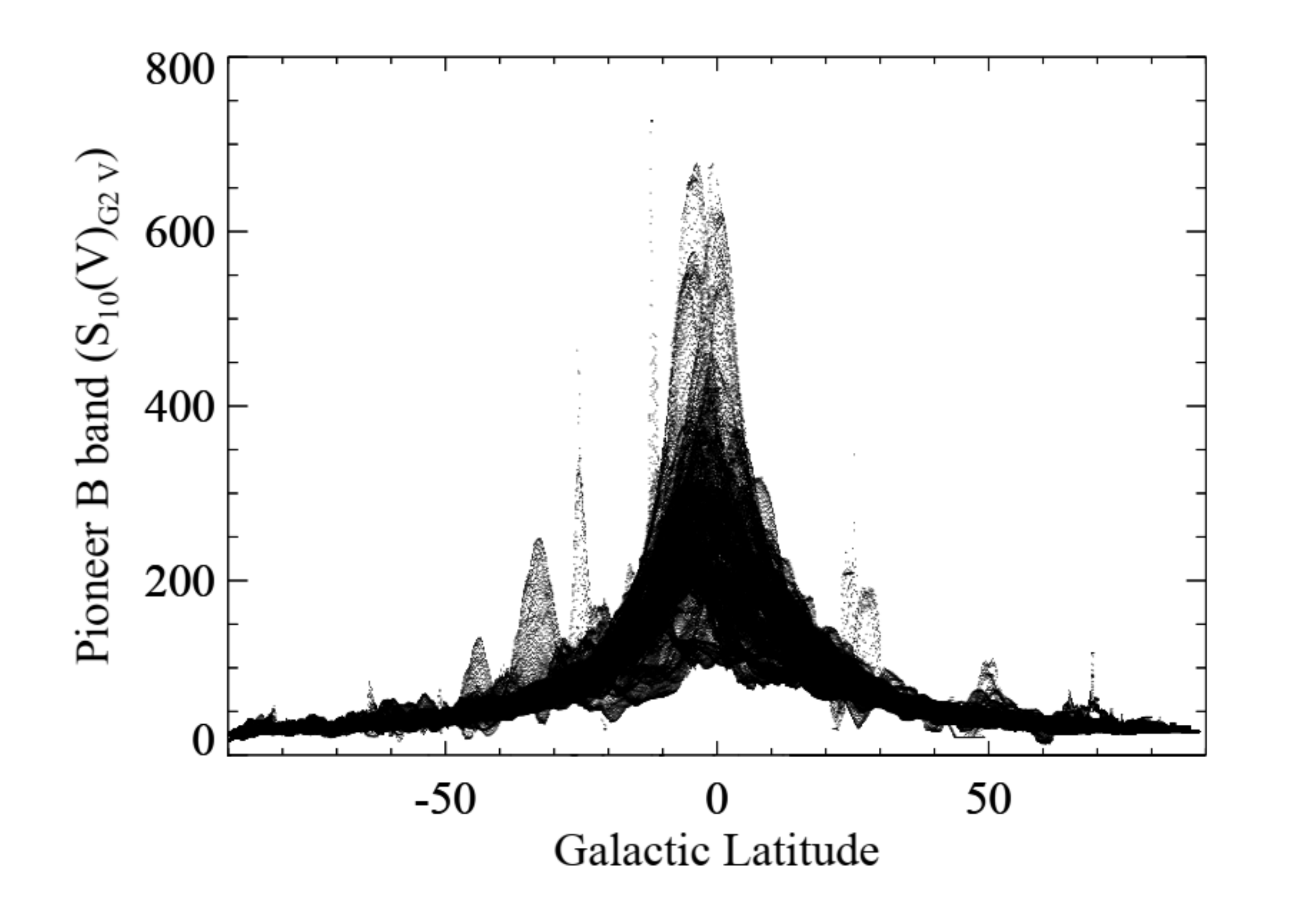} 
 \includegraphics[width=0.45\textwidth]{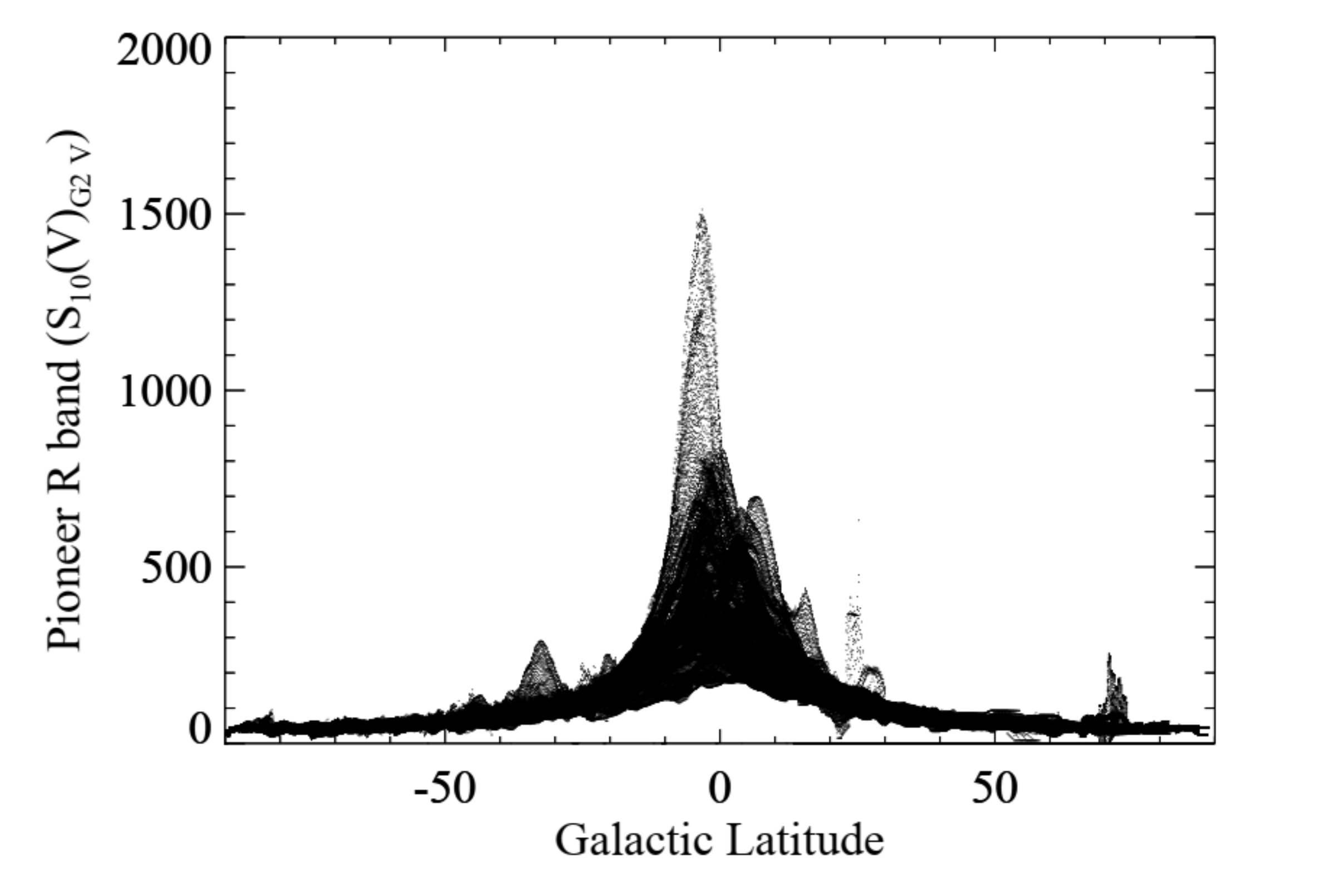} 
 \caption{Clockwise from top left: latitude dependence of FUV background at 1100 \AA\ from \citet{Murthyvoy_all}; latitude dependence of UV background at 1539 \AA\ and 2316 \AA\ from \citet{Murthy2014apj}; and at 4370 and 6441 \AA\ from \citet{Gordon_Pioneer1998}. In each case the cosecant dependence expected of any Galactic emission is apparent but with more scatter at 1100 \AA\ where there are fewer stars and the distances traveled are less because of the optical depth of the interstellar dust.}
  \label{Fig:voy_gb}
\end{center}
\end{figure}

Most of the diffuse radiation that we see in the sky is due to the scattering of starlight from interstellar dust grains \citep{Jura1979}. We have shown the diffuse radiation from {\it Voyager} \citep{Murthyvoy_all} and {\it FUSE} \citep{Murthy_sahnow2004} observations; from \galex FUV and NUV observations \citep{Murthy2014apj}; and {\it Pioneer 10/11} observations \citep{Gordon_Pioneer1998} in Fig. \ref{Fig:euv_bkgd}. Excluding one point from the Large Magellanic Clouds, where there are both stellar photons and dust in abundance, the brightest part of the diffuse sky in any wavelength is the Galactic Plane (Fig. \ref{Fig:voy_gb}) but the distribution of the scattered light is much different in the different bands.

The optical depth of the interstellar dust grains is high in the FUV ($< 1200$ \AA) \citep{Draine_scat2003} and there are few stars hot enough \citep{Murthy_model1995} and bright enough to have a significant number of photons. Hence, there are very few stars which contribute to the FUV sky and the optical depth of the dust is high. Photons therefore do not travel far ($< 600 pc$) and the diffuse light is dominated by the scattering of UV photons from nearby stars by nearby dust grains. Thus there are intense patches such as the Coalsack Nebula near which there are three of the five brightest stars in the UV sky \citep{Murthy_CS1994} while there are dark patches elsewhere in the Galactic Plane, just because there are no nearby stars. 

The optical depth of the dust grains is less at longer wavelengths, particularly into the visible, and there are an increasing number of cool stars which contribute to the interstellar radiation field leading to a smoother distribution of the dust scattered radiation. Note that the Pioneer data are subject to an uncertain subtraction for background stars \citep{Toller_EBL}. 

\begin{figure}[t]
\begin{center}
 \includegraphics[width=0.8\textwidth]{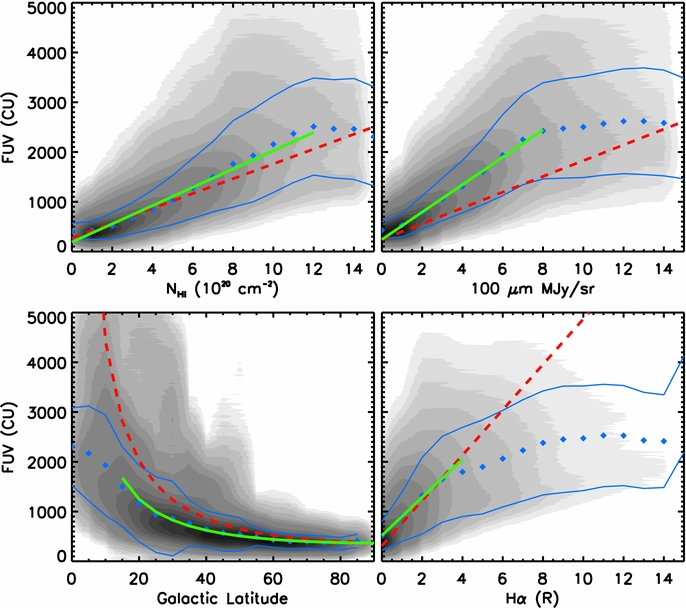} 
 \caption{Two-dimensional histograms of FUV vs N(H I); 100 \micron; Galactic latitude; and H$\alpha$. The red lines are copied from \citet{Seon2011} with the green lines representing the best-fit lines for the median, up to a limit of $1.2 \times 10^{21}$ cm$^{-2}$ and 8 MJy sr$^{-1}$, respectively. The figure is from \citet{Hamden2013} who should be consulted for further details.}
  \label{Fig:uv_ir}
\end{center}
\end{figure}

There is a strong correlation of the UV background with both H I column density and the 100 \micron\ surface brightness (Fig. \ref{Fig:uv_ir}) until the UV flux saturates and the correlation flattens out. This corresponds to an N(H) of about $10^{21}$ cm$^{-2}$ at 1500 \AA. The line of sight becomes more complex at higher column densities and the observed scattering is a convolution of the stellar sources and the scatterers modified by the optical depth. Hence, as noted earlier, there can be dark regions even at low Galactic latitudes. There is a zero-offset which is apparent in Fig. \ref{Fig:voy_gb} but not, interestingly, in the \voyager observations. We will discuss this below.

\subsection{Atomic and molecular emission}

Fluorescent emission in the Lyman band of molecular hydrogen was first discovered by \citet{Hurwitz1994} in the vicinity of the Taurus molecular clouds with Werner band emission observed in the FUV \citep{France2004, france2005, France2005b}. \citet{Jo2017} mapped the Lyman band emission over three quarters of the sky finding that molecular hydrogen averaged about 10\% of the total background in the UV over much of the sky with higher values in high extinction regions where self-shielding allowed the formation of H$_{2}$ molecules.

The first definitive detection of emission lines from the diffuse ISM was from \citet{Murthyvoy1993} who found C III and O VI emission in Eridanus with the first large scale observations attempted by \citet{Welsh_FAUST2007} who mapped hot gas near the North Polar Spur. Although these and the molecular hydrogen emission lines can be important in specific regions of the sky, they are relatively small contributors to the overall diffuse radiation field.

\subsection{Extragalactic Background Light}

The extragalactic background light (EBL) represents the net energy received from the extragalactic sky in any direction. Extending throughout the electromagnetic spectrum, it contains information of the production and redistribution of energy of the universe since the epoch of recombination. The dominant contributor of the EBL in the UV and visible is the integrated galactic light (IGL) which contains the contributions from resolved sources like galaxies and AGNs and is best measured using number counts with galaxy evolution models. The level of the IGL has been estimated to be 150 -- 200 \photu at 1595 \AA\ and 150 -- 215 \photu at 2365 \AA\ \citep{Gardner2000} with lower values of 60 -- 80 \photu and 120 -- 180 \photu at 1530 and 2250 \AA\, respectively, from \citet{Xu2005,Voyer2011,Driver2016,Chiang2019}.

There is relatively little dust scattering from the Galactic poles where there is little dust and the best measurements of the EBL have come from UV observations at the poles. These have found zero-offsets of 200 -- 400 \photu in the poles, of which about half is due to the known IGL. The remaining IGL is, as yet, unknown.

The optical background has been difficult to measure due to the brightness of the foreground \citep{Mattila2017, Mattila_EBL2019} with the best limits on the EBL (200 -- 300 photon units) coming from spacecraft near the edge of the Solar System \citep{Toller_EBL,Zemcov_lorri_2017}.

\section{Future}

\begin{figure}[t]
\begin{center}
 \includegraphics[width=0.8\textwidth]{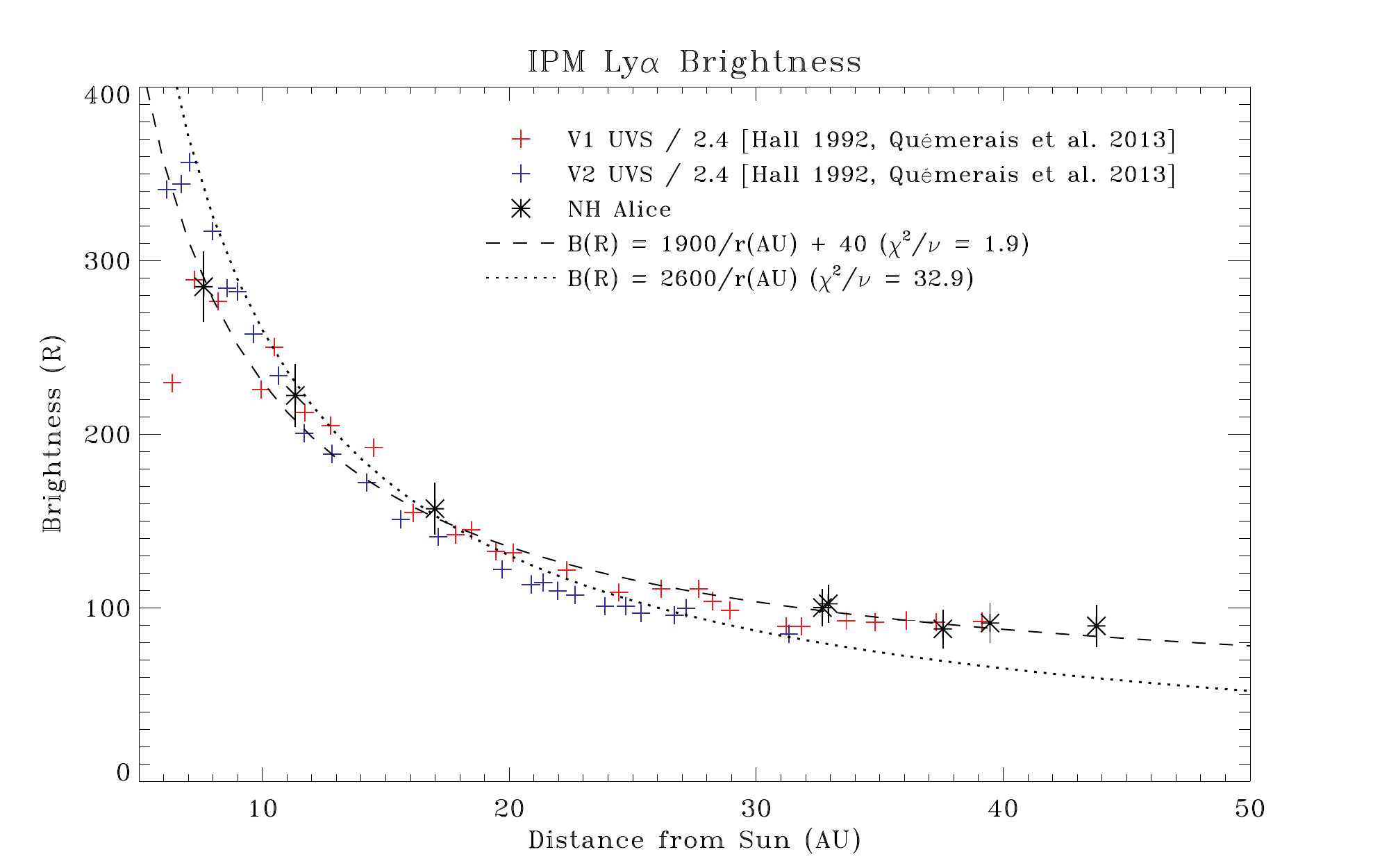} 
 \caption{\lya\ surface brightness as a function of distance from the Sun \citep[private communication]{Gladstone_lya2018}.}
  \label{Fig:lya_fluxes}
\end{center}
\end{figure}

There have been relatively few instruments designed to observe the diffuse radiation field (Table \ref{TAB:missions}) other than a number of sounding rockets which largely observed the North Galactic Pole. The wavelength region from 1400 -- 2000 \AA\ is unique in that there is no interplanetary emission and even a small mission from outside the Earth's atmosphere would be able to probe the Galactic and extragalactic radiation at a high spatial resolution. There are already UV instruments on the Moon \citep{Cao_moon2011} and a small UV imager sent to the Moon with a lander should not be difficult to implement. Observations in other wavelength regions will inevitably be affected by interplanetary emission which will limit the final determination of the astrophysical background. Emission from the interplanetary medium drops rapidly as one leaves the Solar System (Fig. \ref{Fig:lya_fluxes}) and an interstellar probe would, with a relatively small investment in resources, yield a conclusive map of the diffuse background from the UV through the IR \citep{zemcov_nh2018}. 


\section{Acknowledgments}
J.M. acknowledges support from DST/SERB grant EMR/2016/001450. This research has made use of the NASA Astrophysics Data System Bibliographic Services.

\bibliography{murthy}{}
\bibliographystyle{aasjournal}



\end{document}